\documentclass[a4paper,fleqn,usenatbib]{mnras}

\usepackage{newtxtext,newtxmath}

\usepackage[T1]{fontenc}
\usepackage{ae,aecompl}

\usepackage{threeparttable}

\usepackage{color,graphicx}	

\usepackage{amsmath}	
\usepackage{amssymb}	
\usepackage{color}
\usepackage[normalem]{ulem}

\usepackage{etoolbox}
\makeatletter
\patchcmd\@combinedblfloats{\box\@outputbox}{\unvbox\@outputbox}{}{%
   \errmessage{\noexpand\@combinedblfloats could not be patched}%
}%
 \makeatother

\newcommand{\Msun}{\mathrm{M}_{\sun}}
\newcommand{\Mwd}{M_{\textrm{WD}}}
\newcommand{\Rwd}{R_{\textrm{WD}}}
\newcommand{\Mbh}{M_{\textrm{BH}}}
\newcommand{\Enuc}{E_{\textrm{nuc}}}
\newcommand{\rhomax}{\rho_{\textrm{max}}}
\newcommand{\Tmax}{T_{\textrm{max}}}
\newcommand{\rhotmax}{\rho_{\textrm{tmax}}}

\newcommand{\dd}{\textrm{d}}
\newcommand{\secref}[1]{Section~\ref{#1}}
\newcommand{\figref}[1]{Fig.~\ref{#1}}
\newcommand{\tabref}[1]{Table~\ref{#1}}

\title[WD--BH TDEs]{Tidal Disruption of a White Dwarf by a Black Hole:
The Diversity of Nucleosynthesis, Explosion Energy, and the Fate of Debris Streams}

\author[Kojiro Kawana et al.]{
Kojiro Kawana,$^{1}$\thanks{E-mail: kawana@utap.phys.s.u-tokyo.ac.jp}
Ataru Tanikawa,$^{2,3}$
Naoki Yoshida$^{1,4}$
\\
$^{1}$Department of Physics, School of Science, The University of Tokyo, 7-3-1, Bunkyo, Tokyo 113-0033, Japan\\
$^{2}$Department of Earth Science and Astronomy, College of Arts and Sciences, The University of Tokyo, \\
3-8-1 Komaba, Meguro-ku, Tokyo 153-8902, Japan\\
$^{3}$RIKEN Advanced Institute for Computational Science, 7-1-26 Minatojima-minami-machi, Chuo-ku, Kobe, Hyogo 650-0047, Japan\\
$^{4}$Kavli Institute for the Physics and Mathematics of the Universe (WPI), The University of Tokyo, \\
5-1-5, Kashiwanoha, Kashiwa, Chiba 277-8583, Japan
}

\date{Accepted XXX. Received YYY; in original form ZZZ}

\pubyear{2018}

\begin{document}
\label{firstpage}
\pagerange{\pageref{firstpage}--\pageref{lastpage}}
\maketitle

\begin{abstract}
We run a suite of hydrodynamics simulations of tidal disruption events 
(TDEs) of a white dwarf (WD) by a black hole (BH) with a wide range of WD/BH masses and orbital parameters. 
We implement nuclear reactions to study nucleosynthesis and its dynamical effect through release of nuclear energy. 
The released nuclear energy effectively increases the fraction of unbound ejecta.
    This effect is weaker for a heavy WD with 1.2~$\mathrm{M}_{\sun}$, because the
specific orbital energy distribution of the debris is predominantly
determined by the tidal force, rather than by the explosive reactions.
The elemental yield of a TDE depends critically on the initial
composition of a WD, while the BH mass and the orbital parameters
also affect the total amount of synthesized elements.
Tanikawa et al. (2017) find that simulations of WD--BH TDEs with low
resolution suffer from spurious heating and inaccurate nuclear reaction results. 
In order to examine the validity of our calculations, we compare the
amounts of the synthesized elements with the upper limits of them derived in
a way where we can avoid uncertainties due to low resolution. 
The results are largely consistent, and thus support our findings.
We find particular TDEs where early self-intersection of a WD occurs
during the first pericentre passage, promoting formation of
an accretion disc. 
We expect that relativistic jets and/or winds would form in these cases
because accretion rates would be super-Eddington.
The WD--BH TDEs result in a variety of events depending on the WD/BH
mass and pericentre radius of the orbit. 
\end{abstract}


\section{Introduction}
\label{section:intro}
A star passing close to a black hole (BH) can be disrupted
when the tidal force on the star exceeds its self-gravity. 
The disrupted star leaves debris bound
to the BH, but also disperse unbound materials \citep[e.g.][]{1988Natur.333..523R}. 
The bound debris cause emission via various possible
processes (for a review, see \citealt{2018arXiv180110180S}).
One way to produce the emission is collision of advanced and
leading streams of bound debris
\citep{2015ApJ...804...85S,2015ApJ...806..164P,2015ApJ...809..166G,2016MNRAS.455.2253B,2016MNRAS.461.3760H,2016ApJ...830..125J}.
The second way is forming an accretion disc and a flare, where the radiation
could be reprocessed by surrounding materials
\citep{1997ApJ...489..573L,2004ApJ...610..707B,2011MNRAS.410..359L,2013ApJ...767...25G,2014ApJ...783...23G,2016ApJ...827....3R,2016MNRAS.461..948M,2017arXiv170702993R}.
Another way is forming relativistic jets that probably occur when the accretion
rate exceeds the Eddington limit 
(\citealt{2011Sci...333..203B,2011Sci...333..199L,2011Natur.476..421B,2011Natur.476..425Z,2011MNRAS.416.2102G,2012ApJ...753...77C,2014ApJ...781...82C,2015ApJ...809....2C,2015MNRAS.452.4297B,2016ApJ...827..127K};
for a review, see \citealt{2015JHEAp...7..148K}).
In this tidal disruption event (TDE),
we can consider various types of the disrupted star, such as
a main-sequence-type star (MS), giant star, and planet
\citep{2004ApJ...615..855K,2017ApJ...841..132L}.

\begin{figure}
	\includegraphics[width=\columnwidth]{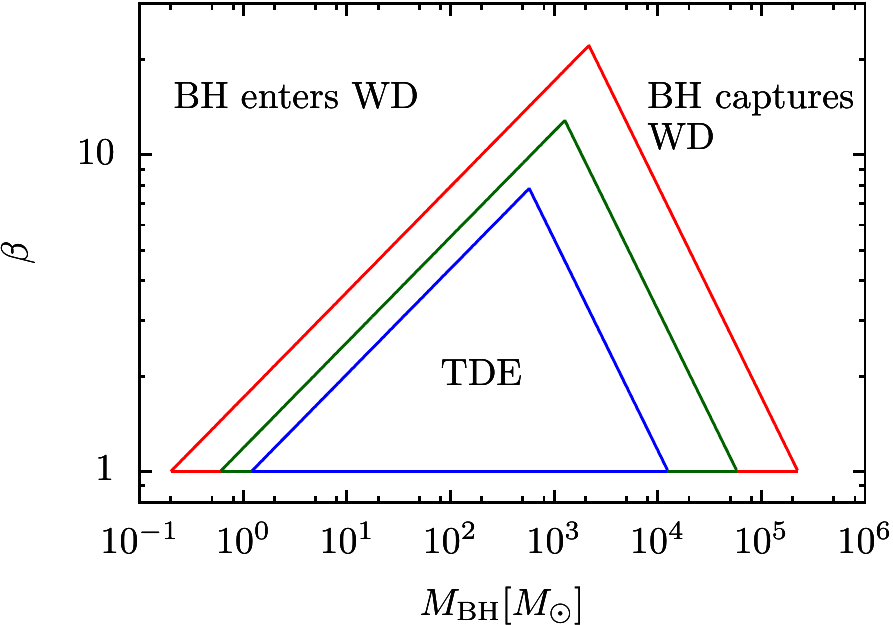}
	\caption{Parameter spaces where WD--BH TDEs occur. 
The red, green, and blue lines are for $\Mwd=0.2$, 0.6, and 1.2\,$\Msun$, respectively.
The areas inside the triangles, respectively, correspond to the parameter spaces where TDEs occur. The original figure is given in \citet{1989A&A...209..103L}
	}
    \label{fig:1}
\end{figure}

TDEs of a white dwarf (WD) have unique features.
First, the range of the BH mass is restricted to stellar and intermediate masses.
A supermassive BH (SMBH) whose mass is larger than $10^5\,\Msun$
does not cause WD--BH TDEs because it swallows a WD before disrupting it tidally;
there would be no observable signals except for gravitational waves
(see Fig.~\ref{fig:1}; 
\citealt{1989A&A...209..103L,2014ApJ...795..135E}). In contrast, other
types of stars, such as MSs
and red giant stars, can be disrupted by an SMBH \citep{2004ApJ...615..855K,2017ApJ...841..132L}. 
Thus, WD--BH TDEs may provide interesting information on the existence and the properties of
intermediate mass BHs (IMBHs). 

The second feature is the explosive thermonuclear reactions of a WD. 
They are caused by strong compression owing to the tidal force perpendicular
to the orbital plane. Such adiabatic compression leads
to shock heating during pericentre passage,
resulting in explosive thermonuclear reactions
\citep{1982Natur.296..211C,1983ApJ...273..749B,1989A&A...209...85L,2008ApJ...679.1385R,2009ApJ...695..404R,2012ApJ...749..117H,
  2015MNRAS.450.4198S,2017ApJ...839...81T}.
If a substantial amount of radioactive nuclei, such as $^{56}$Ni, is synthesized
in unbound ejecta, their decay supplies nuclear energy into the ejecta, 
and the event may appear similar to Type I supernovae (SNe; \citealt{2016ApJ...819....3M}). 
Other signatures from the WD--BH TDEs, such as $\gamma$-ray bursts and gravitational waves,
have also been discussed \citep{2010MNRAS.409L..25Z,2011ApJ...726...34C,2012ApJ...749..117H,2014ApJ...795..135E,2014PhRvD..90f4020C,2015ApJ...804...85S,2016ApJ...833..110I}. 

There are a few key physical quantities that characterize WD--BH TDEs.
One of them is the Schwarzschild radius of the BH $R_S=2 G \Mbh / c^2$, where $G$
is the gravitational constant, $\Mbh$ is the BH mass, and $c$ is the speed of light.
\footnote{Throughout this paper, we assume a BH has no spin.}
Another is the tidal radius $R_t \equiv \Rwd \left( \Mbh / \Mwd \right)^{1/3}$, at which the tidal force exceeds
the self-gravity. It is estimated to be
\begin{equation}
    R_t
\simeq 1.2\times10^{10}
\left (\frac{\Rwd}{10^9\,\textrm{cm}}\right )
\left (\frac{\Mbh}{10^3\,\Msun}\right )^{1/3}
\left (\frac{\Mwd}{0.6\,\Msun}\right )^{-1/3}
\textrm{cm}, 
	\label{eq:tidalradius}
\end{equation}
where $\Rwd$ is the WD radius, and $\Mwd$ is the WD mass. We introduce the dimensionless penetration
parameter $\beta \equiv R_t / R_p$, where $R_p$ is the pericentre radius of the orbit.
A WD is tidally disrupted if $\beta\gtrsim1$. Note that the criterion is slightly modified
depending on the detailed structure of the WD \citep{1989A&A...209..103L,2017ApJ...841..132L,2017A&A...600A.124M}. 
If we adopt general relativity (GR) and a parabolic orbit, there is a relationship between the pericentre radius $R_p$
and the specific angular momentum $j$ (see e.g. \citet{1983bhwd.book.....S}),
\begin{equation}
    j= (R_S R_p c^2)^{1/2} \left(1 - \frac{R_S}{R_p}\right)^{-1/2}.
	\label{eq:j}
\end{equation}
The boundary on whether a BH swallows a WD is expressed as $j =2 R_S c$, $R_p = 2 \, R_S$, or 
\begin{equation}
	\beta \simeq 10
\left (\frac{\Rwd}{10^9\,\textrm{cm}}\right )
\left (\frac{\Mbh}{10^3\,\Msun}\right )^{-2/3}
\left (\frac{\Mwd}{0.6\,\Msun}\right )^{-1/3}
.
	\label{eq:beta1}
\end{equation}
If $j <2 R_S c$, Equation~\ref{eq:j} does not have a real solution for $R_p$.
This corresponds to the situation where the WD would be directly captured by the BH,
and then there would be no observable signals except for gravitational waves. 
The BH can also swallow a part of the WD in the first pericentre passage if $R_p \leq \Rwd$, or
\begin{equation}
	\beta \gtrsim12
\left (\frac{\Mbh}{10^3\,\Msun}\right )^{1/3}
\left (\frac{\Mwd}{0.6\,\Msun}\right )^{-1/3}
.
	\label{eq:beta2}
\end{equation}
This situation is conventionally called, ``the BH enters the WD" \citep{1989A&A...209..103L}.
Note that in that case the BH is much smaller than the WD so that a tiny part of the WD is swallowed during the pericentre passage. 
As we will discuss in \secref{section:results}, an interesting class of TDEs occur in this case,
where the WD is so strongly compressed that nuclear burning is triggered and the foregoing part of the disrupted debris hits the trailing part. 

The parameter space where WD--BH TDEs occur is shown in Fig.~\ref{fig:1}.
Thus, an SMBH with $\Mbh\gtrsim10^5\,\Msun$ cannot tidally disrupt a WD. 
This feature of WD--BH TDEs is useful to investigate the properties of IMBHs.
Once many WD--BH TDEs are observed, the number density and 
growing process of IMBHs could be constrained. 
So far, only a few candidates of WD--BH TDEs have been reported \citep{2011ApJ...743..134K,2013ApJ...769...85S,2013ApJ...779...14J,2016ApJ...833..110I,2017MNRAS.467.4841B}. Next-generation transient
surveys, such as the Large Synoptic Survey Telescope, may detect
a few tens of the events \citep{2016ApJ...819....3M}. It is expected that there would be
a variety of observational signals originating from the diversity of the TDEs.
Therefore, theoretical templates for various types of WD--BH TDEs are needed
in order to detect them robustly by the future transient surveys.

If a WD passes very close to the BH, not only tidal disruption but also explosive thermonuclear
burning is likely ignited by adiabatic compression and shock heating
\citep{1982Natur.296..211C,1983ApJ...273..749B,1989A&A...209...85L,2009ApJ...695..404R,2017ApJ...839...81T}.
\citet{2009ApJ...695..404R} use numerical simulations to study the WD--BH TDEs for 16 parameter
sets. They conclude that the critical condition of the explosive thermonuclear reactions
is $\beta \gtrsim3$ for $1.2\,\Msun$ WDs. They also show that the release of nuclear energy increases the fraction
of the unbound ejecta from about 50\% to 65\% of the initial mass of the WD, which affects the
flare caused by the accretion of the bound debris on to the BH \citep{2014ApJ...794....9M}.
\citet{2016ApJ...819....3M} study observational signatures and estimate a detection rate of the events
using the results of \citet{2009ApJ...695..404R}. They argue 
that the WD--BH TDEs with the explosive nuclear reactions would be the reminiscent of Type I SNe.

The 16 parameter sets in \citet{2009ApJ...695..404R} are not systematically
chosen.
They are not enough to comprehensively explore the
parameter spaces shown in \figref{fig:1}.
The problem makes it difficult to reveal the variety of WD--BH TDEs
and to determine detailed critical conditions on explosive nuclear reactions for various types of WDs.

In this paper, we explore the variety of WD--BH TDEs, 
focusing on nucleosynthesis and its effects on the debris of the disrupted WD.
We perform a systematic and comprehensive parameter study for 184
parameter sets, 
changing $\Mwd$, $\Mbh$, and $\beta$. 
To this end, we use smoothed particle hydrodynamics (SPH) simulations
coupled with the nuclear reactions of $\alpha$-chain networks among the 13 species from $^4$He to $^{56}$Ni. 
Our numerical methods are largely based on the recent study of
\citet{2017ApJ...839...81T}.

\citet{2017ApJ...839...81T} show that the results of nucleosynthesis
do not converge even with 25 millions SPH particles.
They estimate that, in order to follow shock heating and ignition of
the nuclear reactions and detonations correctly,
structures of very small spatial scales ($\lesssim10^6$\,cm) must be resolved. 
In order to satisfy this condition, we must pay extremely expensive computational costs, 
such as $\gg 10^9$ SPH particles.
In addition, we must perform many simulations
with various parameter sets in order to perform parameter study. 
This point also makes it difficult for us to perform simulations with
very high resolution.

For these reasons, we perform simulations with inadequate resolution
failing to resolve the shock heating.
Our results on nucleosynthesis have uncertainties due to the low
resolution.
However, we fix the number of SPH particles in all runs, which enables us
to explore the dependence on the parameters based on homogeneous samples.
We also examine the validity of our calculations by comparing amounts of
synthesized elements with upper limits of them given in a different
way where we can avoid the uncertainties due to low resolution.

The structure of the rest of the present paper is as follows. In
\secref{section:methods}, we describe methods and setups for our
numerical simulations. In Section~\ref{section:results}, we show results
of our simulations and discuss them.
In Section~\ref{section:resolution}, we discuss the resolution
dependence of our results and examine the validity of our results.
In Section~\ref{section:conclusions}, the summary of this paper and the implications are given.
\section{Numerical Methods}
\label{section:methods}
We follow the WD--BH TDEs with a wide variety of configurations by means of SPH simulations
coupled with the nuclear reactions. We largely follow \citet{2017ApJ...839...81T}.
Here, we summarize the methods and differences from them. 

The key simulation parameters are $\Mwd$, $\Mbh$, and $\beta$,
with the respective range
from $[0.2\,\Msun,\,1.2\,\Msun]$, $[10\,\Msun,\,10^{5}\,\Msun]$, and $[1,\, 5.5]$.
We model a WD as a collection of SPH particles and a BH as a single gravity source.
We adopt the popular SPH equations using the Wendland $C^2$
kernel \citep{wendland1995piecewise,2012MNRAS.425.1068D} for the interpolation of SPH kernels.
We take the artificial viscosity parameters suggested by \citet{1997JCoPh.136..298M}. 
We use FDPS \citep{2016ascl.soft04011I,2016PASJ...68...54I} and explicit AVX
instructions \citep{2012NewA...17...82T,2013NewA...19...74T} to achieve fast
calculations on distributed-memory parallel supercomputers.

We adopt the Helmholtz equation of state (EoS; \citealt{2000ApJS..126..501T}) for the WD.
The EoS considers, with Coulomb corrections, degenerate electron/positron gas, ion gas as an ideal gas
with the adiabatic index $\gamma=5/3$, and radiation pressure of photons.
We incorporate the nuclear reactions using Aprox13 \citep{2000ApJS..129..377T}.
This covers $\alpha$-chain reaction networks of the 13 nuclear species from $^4$He to $^{56}$Ni. 
The networks can give the energy generation rate within
$\sim30$\% errors compared with much larger networks, which enables us
to reasonably track the abundance levels.
In order to calculate the Helmholtz EoS and Aprox13, we use the routines developed
by the Center for Astrophysical Thermonuclear Flashes at the University of Chicago.

The gravitational potential of the BH is represented by the generalized Newtonian
potential proposed by \citet{2013MNRAS.433.1930T}, which is an excellent approximation for
a Schwarzschild BH. 
We remove SPH particles when a distance between a SPH particle and the centre of the BH is smaller than the sum of the Schwarzschild radius and the kernel support radius of the SPH particle. 
We implement the self-gravity of the WD with with adaptive gravitational softening \citep{2007MNRAS.374.1347P}. 

We consider three types of WDs (see Table~\ref{tab:Initwd}). 
We assume that the chemical
compositions of the WDs are homogeneous, following the ways of
\citet{2009ApJ...695..404R} and \citet{2017ApJ...839...81T}.
Note that the inhomogeneity play important roles if it is realized. 
\citet{2017ApJ...841..132L} perform hydrodynamical simulations of TDEs
of a He WD with a tenuous hydrogen envelope.
They show that the TDEs have unique fallback rates. 
\citet{2017arXiv171107115T} shows that in TDEs of a CO WD with a sufficient
amount of He envelope ($\gtrsim 5 \, \%$ of the total WD mass), the He
envelope first detonates that drives CO core detonation as well. 
We generate and relax initial distributions of the SPH particles
in the same manner as in \citet{2015ApJ...807...40T}, \citet{2015ApJ...807..105S}, and \citet{2016ApJ...821...67S}. 
We employ 786,432 SPH particles to represent a WD.
We discuss resolution dependence of our results in
\secref{section:resolution}.

For each run, the initial orbital parameters are set such that the
orbit should be parabolic in the Schwarzschild metric. 
The initial separation between the BH and the WD is set to be $5 \,R_t$. 
This enables us to take into account a small tidal deformation of the WD before the distance between the WD and the BH becomes the tidal radius.
We take the termination time as the twice of the time when the WD passes the pericentre in Newtonian gravity.
The nuclear reactions nearly cease at the end of the simulations. 
In most cases, bound debris of the disrupted WD have not yet been swallowed by the BH at the end. 
However, in the other cases where the BH and the WD encounter very
closely (Type III TDEs explained in Section.~\ref{section:results}), a small part of the WD is swallowed by the BH at the end.
\begin{table}
	\centering
	\caption{Initial conditions of WDs. We show $\Mwd$ in units of $\Msun$, $\Rwd$ in $10^8$\,cm, and $\rho_c$ in g\,cm$^{-3}$, which is the central density of a WD.}
	\label{tab:Initwd}
	\begin{tabular}{cccc} 
		\hline
		$\Mwd$ & $\Rwd$& $\rho_c$& Compositions \\
		\hline
		0.2 & 13 & $2.2\times10^5$ & $^{4}$He 100\% \\
		0.6 & 7.5 & $3.6\times10^6$ & $^{12}$C 50\% $^{16}$O 50\% \\
		1.2 & 3.4 & $1.5\times10^8$ & $^{16}$O 60\% $^{20}$Ne 35\% $^{24}$Mg 5\% \\
		\hline
	\end{tabular}
\end{table}
\section{Results and discussion}
\label{section:results}
\begin{figure*}
\centering
\includegraphics[width=0.98\hsize]{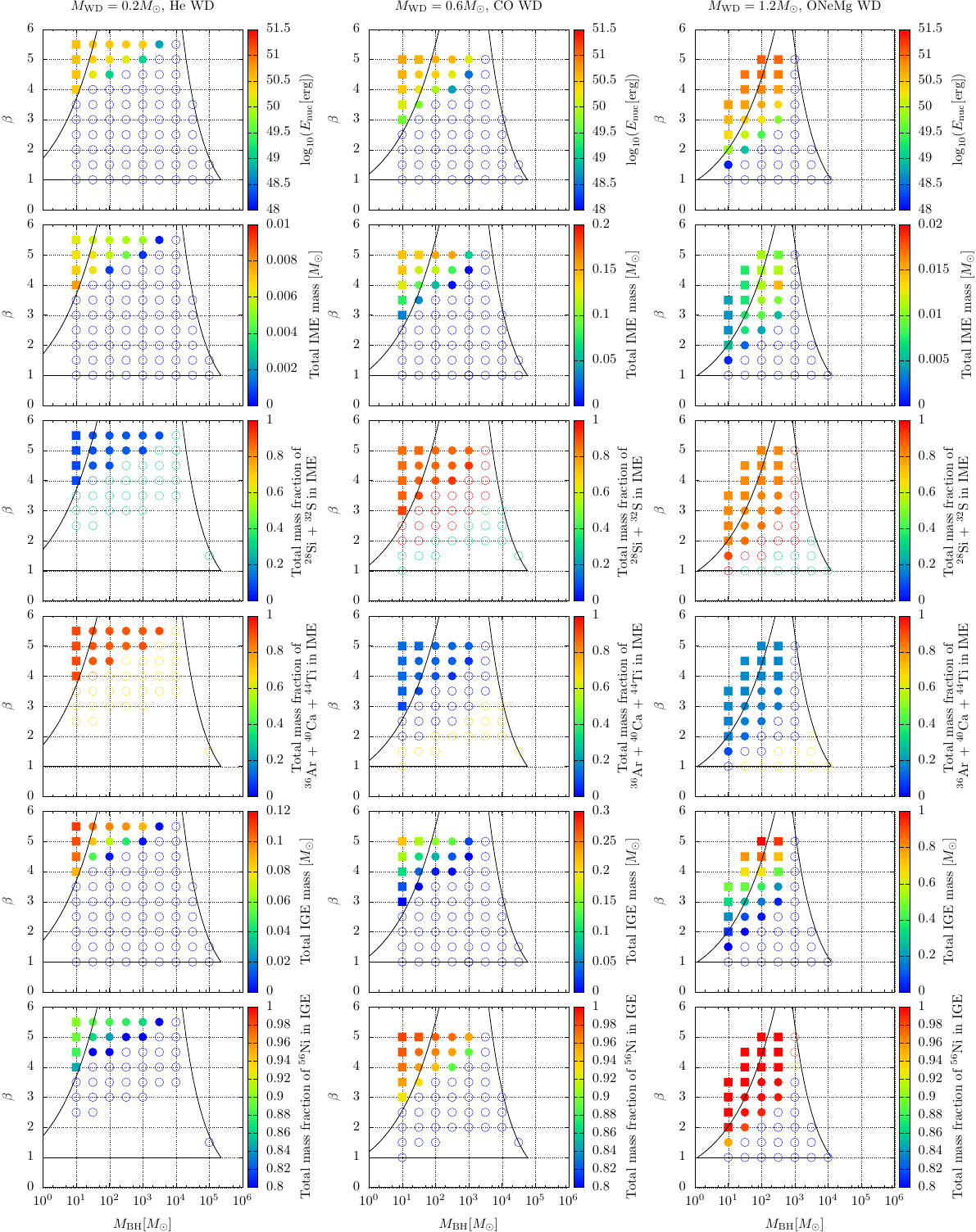}
\caption{
  Properties of the TDE debris. The horizontal and vertical axes are the same
  as Fig.~\ref{fig:1}, but with the latter in linear scale here.
  The solid curves also correspond to those in Fig.~\ref{fig:1}.
  From left to right, each column shows, respectively, the results for $\Mwd=0.2$, 0.6, or 1.2\,$\Msun$. The open circles show TDEs without explosive nuclear reactions (Type I), the filled circles are TDEs with the explosive nuclear reactions but without early self-intersection (Type II). The filled squares show TDEs with the explosive nuclear reactions and self-intersection (Type III).
}
\label{fig:fig_GN_tot_show}
\end{figure*}
\begin{figure*}
\centering
\includegraphics[width=\hsize]{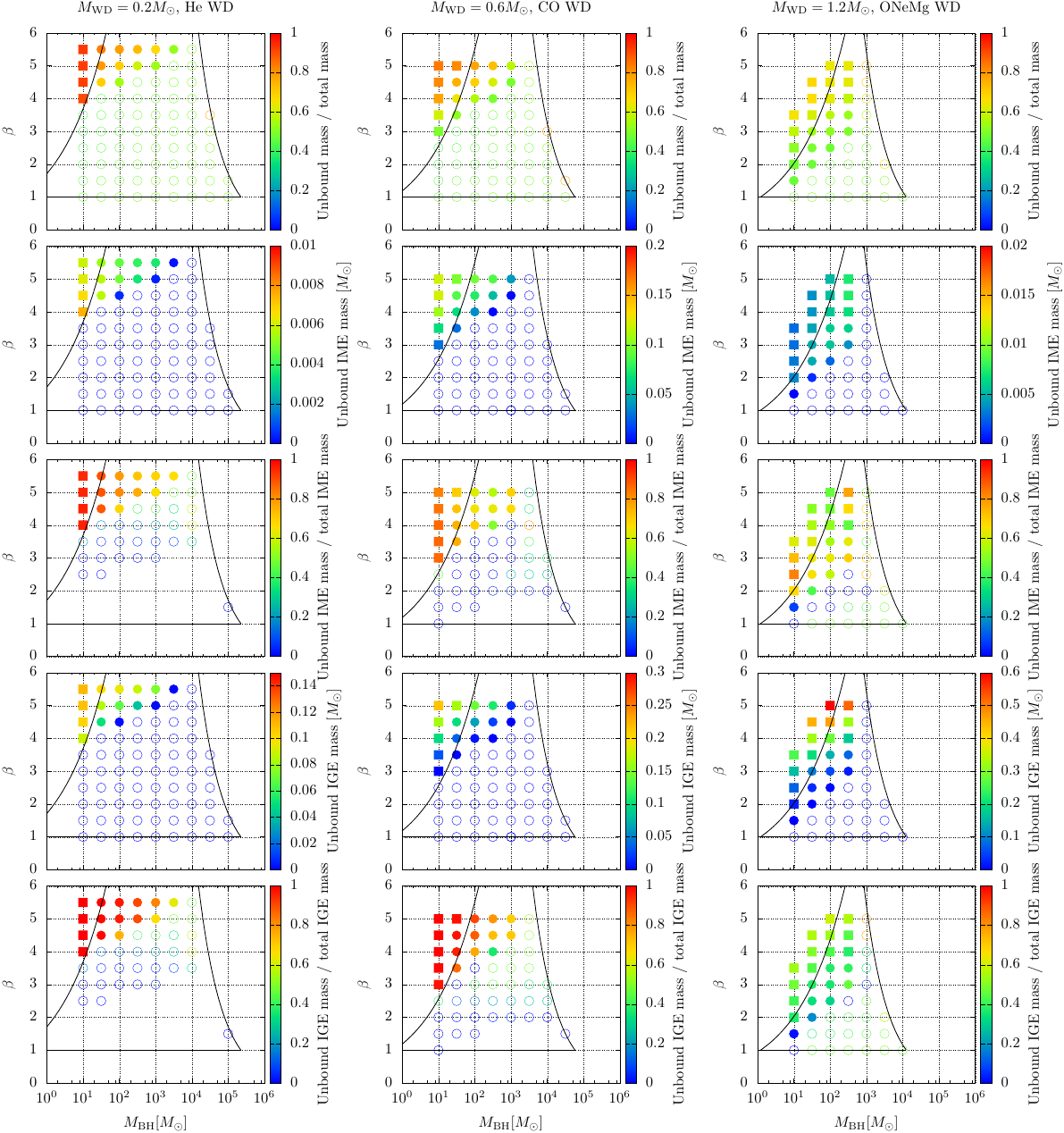}
\caption{
Properties of unbound ejecta. We show the ratio of unbound ejecta mass to the total debris mass, unbound IMEs/IGEs mass, and ratio of unbound IMEs/IGEs mass to the total IMEs/IGEs mass.
}
\label{fig:fig_GN_unbound_show}
\end{figure*}
\subsection{A variety of WD--BH TDEs}
Fig.~\ref{fig:fig_GN_tot_show} shows the physical properties of total (bound $+$ unbound)
debris, while Fig.~\ref{fig:fig_GN_unbound_show} shows those of unbound debris.
The first row in Fig.~\ref{fig:fig_GN_tot_show} shows the released nuclear energy $\Enuc$.
With $\Enuc\lesssim10^{48}$~erg, the nuclear reactions are unimportant,
and thus we take this value as the threshold of whether the explosive nuclear reactions occur.
Then, our results can be categorized into the following three types. 

In Type I TDEs, a WD is tidally disrupted, but explosive nuclear
reactions are not ignited.
We show the corresponding cases with the open circles
in Figs~\ref{fig:fig_GN_tot_show} and \ref{fig:fig_GN_unbound_show}. 
This type can be considered as an analogue to ordinary TDEs where an MS
is disrupted without triggering explosive nuclear reactions.

\begin{figure}
\centering
\includegraphics[width=\hsize]{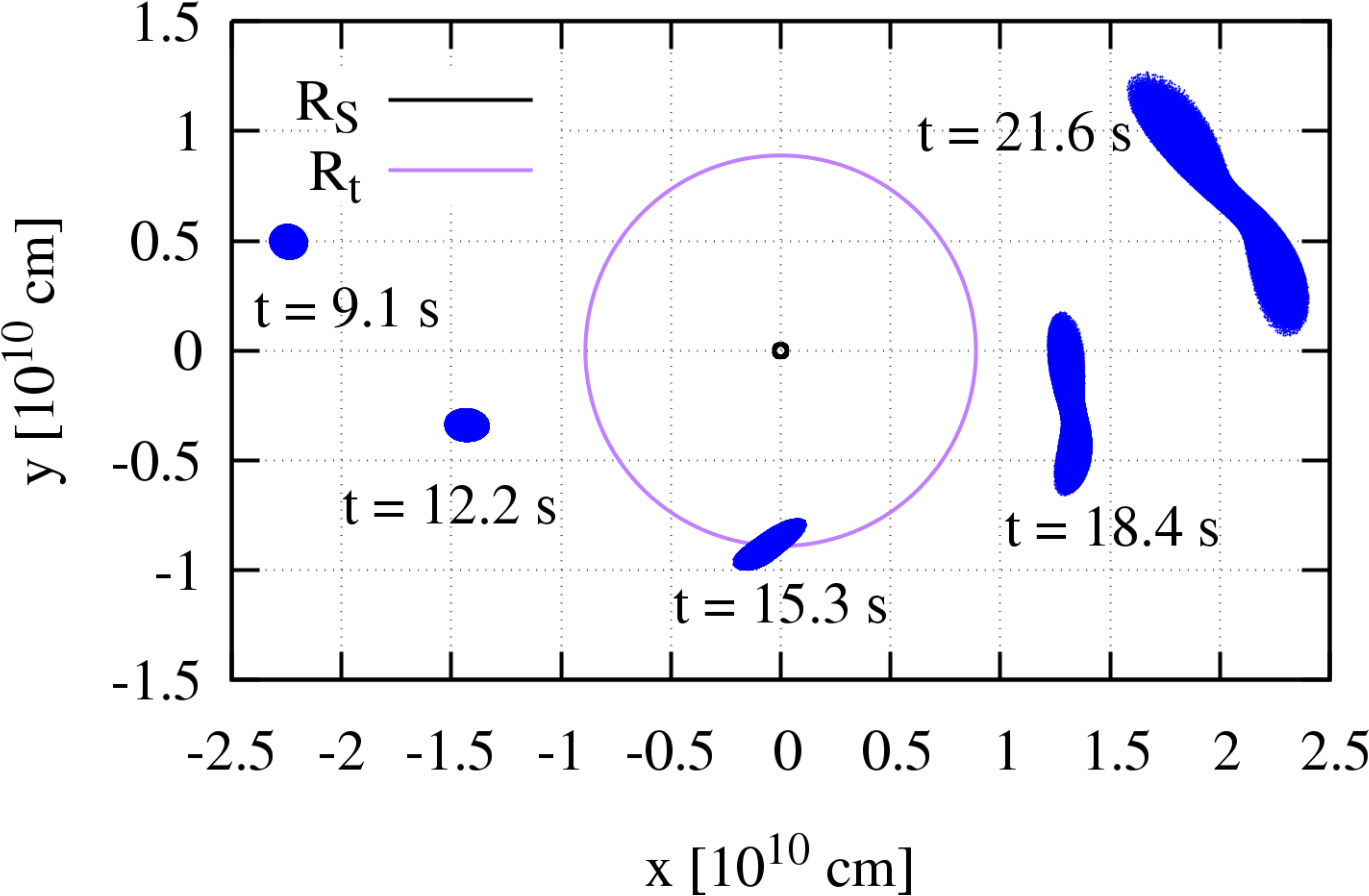}
\caption{A Type I TDE, where explosive nuclear reactions do not occur. 
The parameters are $\Mwd=0.6\,\Msun,\, \Mbh=10^{3}\,\Msun$ and $\beta=1.0$. }
\label{fig:fig_normal}
\end{figure}

\begin{figure}
\centering
\includegraphics[width=\hsize]{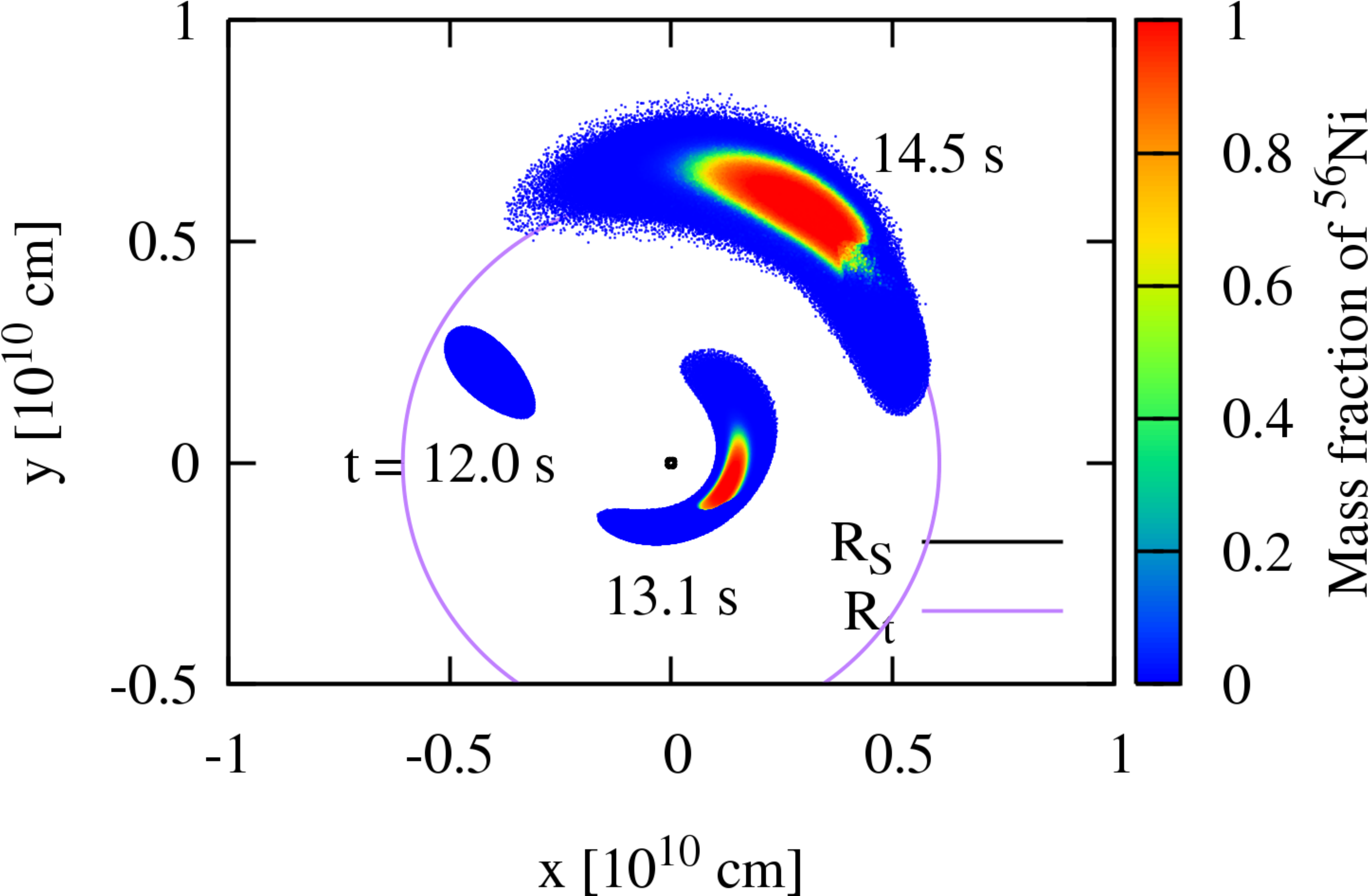}
\caption{A Type II TDE, where explosive nuclear reactions occur without early self-intersection. 
The parameters are $\Mwd=0.6\,\Msun,\, \Mbh=10^{2.5}\,\Msun$, and $\beta=5.0$.}
\label{fig:fig_burn_lowres}
\end{figure}

The other two types show clear signatures of explosive nuclear reactions.
The filled points in Figs~\ref{fig:fig_GN_tot_show} and \ref{fig:fig_GN_unbound_show} represent these
types. The difference between the two types is whether early self-intersection occurs.
In Type II TDEs, the disrupted WD debris does not hit itself, whereas
in Type III, the foregoing part of the debris hits the trailing part.
We mark Type II cases, where the self-intersection is not found, with the filled circles
in Figs~\ref{fig:fig_GN_tot_show} and \ref{fig:fig_GN_unbound_show}.
We show a typical Type II TDE in Fig.~\ref{fig:fig_burn_lowres}.
Type III TDEs are marked with filled squares in Figs~\ref{fig:fig_GN_tot_show} and \ref{fig:fig_GN_unbound_show}. 
Although most Type III TDEs are located in the parameter space that is
conventionally considered as ``a BH enters a WD'', a very small fraction of the WD is swallowed by the BH during the pericentre passage because the BH is much smaller than the WD. 

\begin{figure*}
\centering
\includegraphics[width=\hsize]{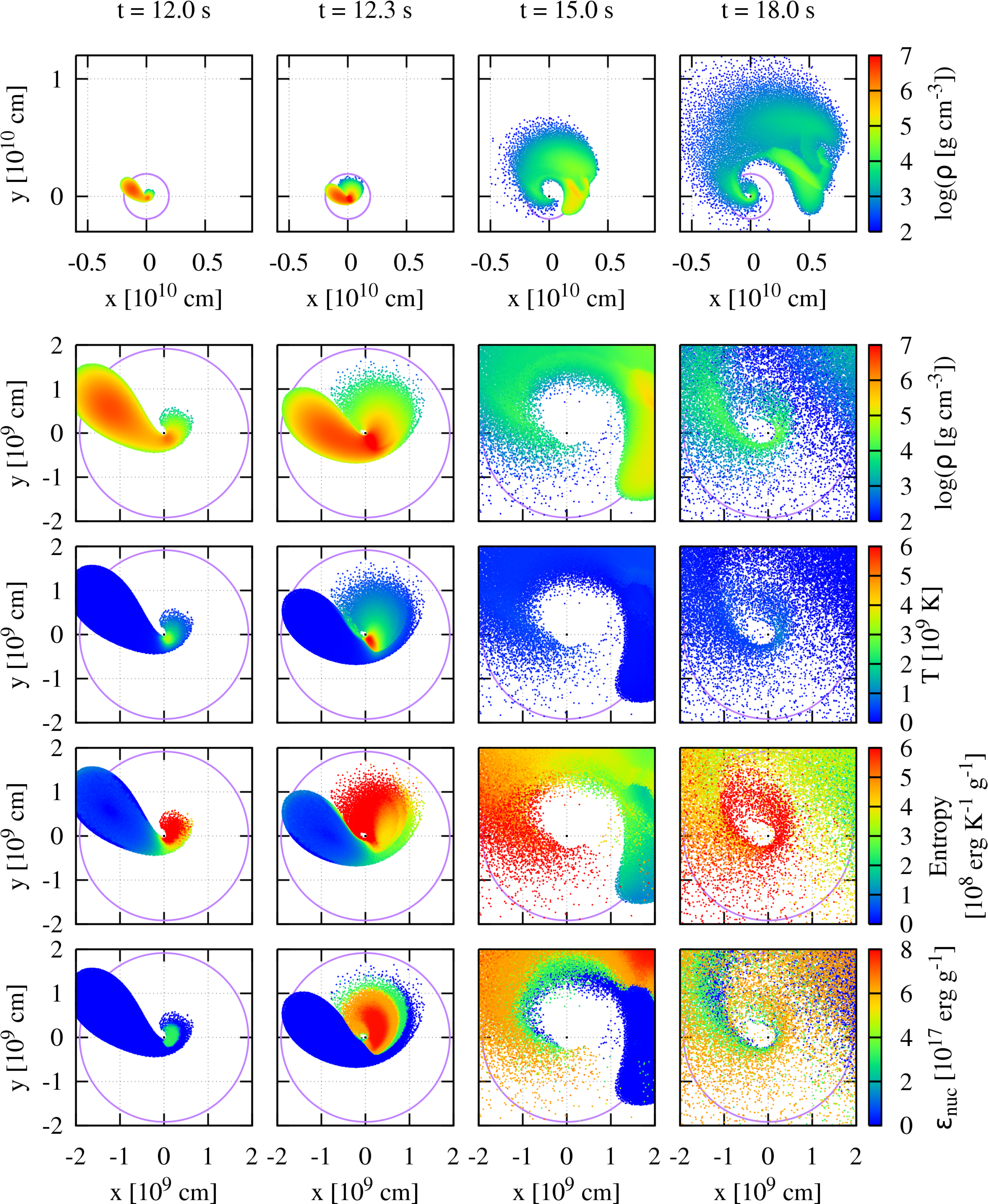}
\caption{%
A Type III TDE, where explosive nuclear reactions occur with early self-intersection. 
The parameters are $\Mwd=0.6\,\Msun,\, \Mbh=10\,\Msun$, and $\beta=5.0$. 
$t =12.0$ and $12.3$~s are respectively just before and after the self-intersection. 
The fifth column shows the released specific nuclear energy. 
The black and purple circles show the Schwarzschild radius and tidal radii respectively. 
Note that the Schwarzschild radius is very small and hard to see in the figure. 
}
\label{fig:fig_self_intersection_unify}
\end{figure*}

We show an example of Type III in Fig.~\ref{fig:fig_self_intersection_unify}. 
The left two columns show effects of the self-intersection. 
The trailing part of the WD intersecting with the advanced part
increases the entropy and temperature up to $\sim2\times10^9$~K,
but nuclear reactions are not ignited. 
The failure of the self-intersection to ignite nuclear reactions is
common among all our runs.
Thus, the effects of the self-intersection on nuclear reactions are
unimportant.

Note that our estimate of the effects of the self-intersection are quantitatively
uncertain due to the following two reasons. 
One reason is that the edges intersecting with each other 
are under-resolved in SPH methods.
The second reason is that SPH particles passing very close to the BH are
removed in our methods, which artificially suppresses the innermost self-intersection.
This self-intersection would be more violent than what we see currently
because bound debris passing closer to the BH have more kinetic energy than 
debris passing farther.
If it is taken into account, the shock would be stronger and
might possibly ignite nuclear reactions.

\begin{figure}
\centering
\includegraphics[width=\hsize]{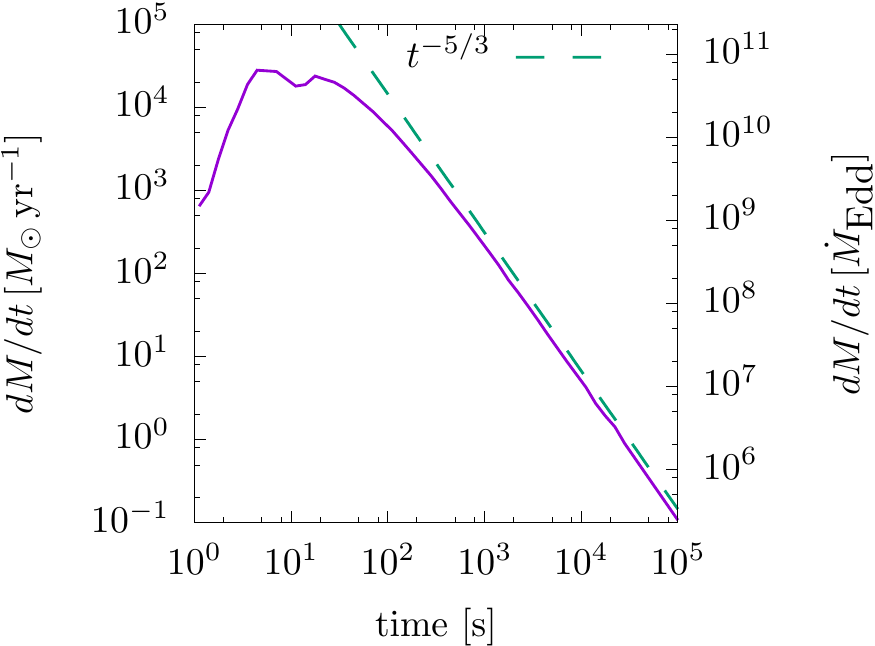}
\caption{%
    A fallback rate of bound debris for the same case as shown in
    \figref{fig:fig_self_intersection_unify}.
}
\label{fig:mdot}
\end{figure}

The early self-intersection circularizes the bound debris, which
promotes formation of an accretion disc.
In Type III TDEs, fallback time-scale of the bound debris is short because the BH masses are small.
In \figref{fig:mdot}, we show the fallback rate calculated from the
distribution of specific orbital energy, $\dd M / \dd \epsilon$, at the end of
the simulation.
Here, we take the radiation efficiency as 0.1 and the electron fraction
$Y_{\mathrm{e}}$ as 0.5.
The peak fallback rate extremely exceeds the Eddington accretion rate.
The early self-intersection enables the fallback debris
to rapidly circularize and to form an accretion disc 
(\citealt{2014ApJ...783...23G,2015ApJ...804...85S,2015ApJ...809..166G,2016MNRAS.455.2253B};
for a recent review, see \citealt{2018arXiv180110180S}).
Thus, we expect that the accretion rate would follow the fallback rate
and would exceed super-Eddington accretion rate.

\citet{2015ApJ...805L..19E} study a similar kind of TDEs to the Type III
TDEs. 
They simulate TDEs where an MS
encounters with an IMBH with $\Mbh=10^5\,\Msun$ very closely ($\beta=10$
and $15$) although these parameter sets still locate in the ``TDE'' region
in the parameter space shown in \figref{fig:1}.
Note that the boundaries shown in \figref{fig:1} are not for a
MS but for WDs.
Common properties between their cases and the Type III TDEs are as follows:
(1) a very close encounter of a star with a BH, 
(2) prompt formation of an accretion disc, and
(3) highly super-Eddington accretion.
A unique point of their cases is that the accretion rates do not
follow canonical $t^{-5/3}$ law but remain
roughly constant for a few days.
They conclude that this is likely due to the strong GR effects. 

Despite the similarities, 
we expect that the accretion rates for the Type III TDEs follow
$t^{-5/3}$ law inferred from \figref{fig:mdot}.
This is because the GR effects are unimportant for the Type III TDEs.
Because the BH masses are small in the Type III TDEs, the ratio of Schwarzschild radii to the pericentre
radii are much smaller than unity.
This point shows that the GR effects are unimportant
for the Type III TDEs, and the canonical accretion rates proportional to $t^{-5/3
}$ would appear. 

\subsection{Abundances}
We further study the TDEs with explosive
nuclear reactions in detail i.e., the filled points in Fig.~\ref{fig:fig_GN_tot_show}. 
The second to fifth rows in Fig.~\ref{fig:fig_GN_tot_show} show masses and fractions of
nuclear species synthesized. All the debris materials are included, irrespective of bound or unbound. 
The explosive burning yields a large amount of IGEs, of which more than $80\%$ is $^{56}$Ni. 
It is expected that the abundant $^{56}$Ni in the unbound debris
can be a power source in late phases of the TDEs. 
We will discuss this possibility in Section~\ref{section:conclusions}. 
The elemental abundance depends on $\Mwd$,
but only weakly on $\Mbh$ and $\beta$. 
Note that we assume different initial compositions of WDs with different $\Mwd$.
Hence the variety of the final elemental abundance
also reflects the initial compositions. 
The total amount of synthesized elements depends not only on $\Mwd$ but
also on $\Mbh$ and $\beta$.

\begin{figure}
\centering
    \includegraphics[width=\hsize]{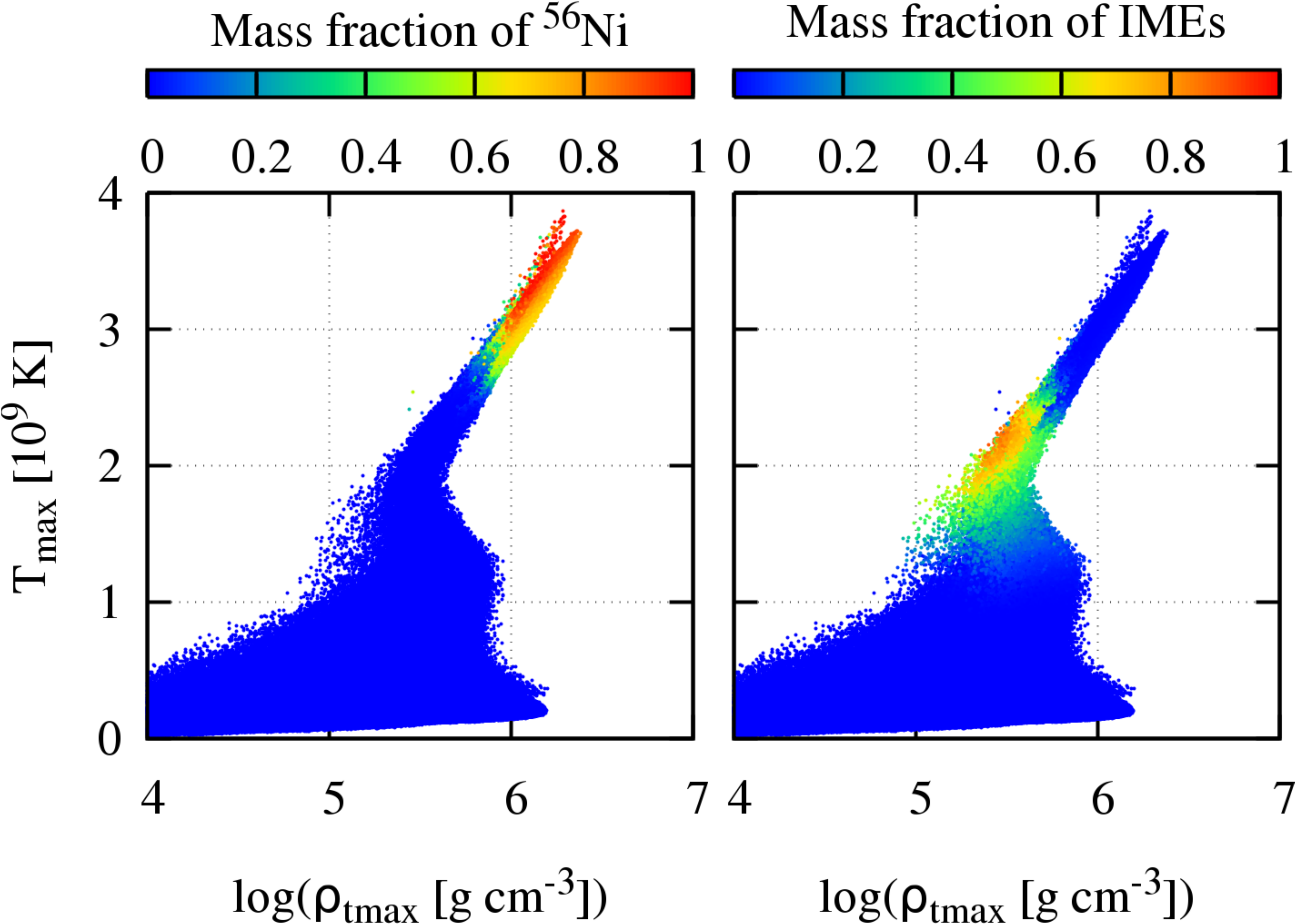}
\caption{%
Mass fraction of $^{56}$Ni and IMEs as functions of the maximum temperature $T_{\textrm{max}}$ of a particle and the density at the moment $\rhotmax$. 
    This is the case of $\Mwd=0.2\,\Msun,\, \Mbh=10^{2.5}\,\Msun$, and $\beta=5.0$.
}
\label{fig:TRhotmax}
\end{figure}

Let us first focus on $\Mwd=0.2\,\Msun$ cases, where a WD is assumed to
consist of pure $^4$He initially. 
These cases typically produce, (1)
a small amount of IMEs, $\lesssim0.01\,\Msun$,
(2) the IMEs mainly consisting of $^{36}$Ar, $^{40}$Ca, and $^{44}$Ti, and
(3) a relatively low fraction of $^{56}$Ni, $\lesssim90\%$ of the IGEs.
These features are qualitatively consistent
with previous works on nucleosynthesis caused by detonations in light He WDs
\citep{1994ApJ...423..371W,1995ApJ...452...62L,2013ApJ...771...14H}. 
\citet{2013ApJ...771...14H} study conditions for detonations in light WDs
using one-dimensional (1D) hydrodynamics simulations. 
They find that, if detonation occurs with $\rhotmax\lesssim10^6$~g~cm$^{-3}$,
where $\rhotmax$ is the density at a temperature maximum,
the dominant nucleosynthesis products are $^{40}$Ca, $^{44}$Ti, and $^{48}$Cr. 
Interestingly, we find that the main product from the nucleosynthesis
is still $^{56}$Ni, and that a very small amount of IMEs left with $\lesssim0.01\,\Msun$. 
This is because the part of the disrupted WD that experiences nuclear reactions
have densities $\rhotmax\gtrsim10^6$~g~cm$^{-3}$ in our simulations.
Then the main product is $^{56}$Ni (see Fig.~\ref{fig:TRhotmax}). 

Our simulations for $\Mwd=0.6$ and $1.2\,\Msun$ show
that $^{28}$Si and $^{32}$S are main IMEs 
and that most of the IGEs is $^{56}$Ni ($\gtrsim95\%$).
A notable difference between the two cases is the synthesized amount of IMEs.
Roughly $0.1\,\Msun$ IMEs are synthesized for $\Mwd=0.6\,\Msun$ cases,
because a significant fraction of the debris experiences incomplete nuclear
reactions. For $\Mwd=1.2\,\Msun$ cases, nuclear reactions proceed completely
and only a small amount ($\simeq0.01\,\Msun$) of IMEs are synthesized.

\subsection{Effects of nuclear reactions on motions of debris}
\label{subsection:unbound}
The first row in Fig.~\ref{fig:fig_GN_unbound_show} shows the ratio of unbound
ejecta mass to total debris mass. 
As is naively expected, the unbound mass is larger for larger $\Enuc$. 
A part of the released nuclear energy
is converted into orbital energy of the debris. 
However, for $\Mwd=1.2\,\Msun$ cases, the fraction of unbound ejecta does
not increase as high as in the cases with lighter WDs.
The fifth row in Fig.~\ref{fig:fig_GN_unbound_show} also shows that
the nucleosynthesis products are mostly unbound
for the $\Mwd=0.2$ and 0.6\,$\Msun$ cases,
while a significant fraction is bound for the $\Mwd=1.2\,\Msun$ cases. 

\begin{figure}
	\includegraphics[width=\hsize]{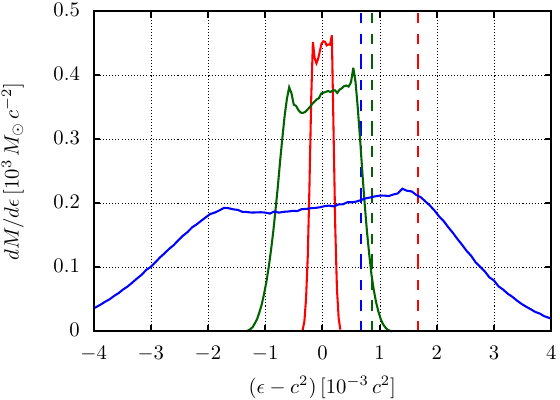}
	\caption{Specific orbital energy distributions after TDEs without explosive nuclear reactions. We compare three cases where $\Mwd=0.2$ (red), 0.6 (green), and 1.2\,$\Msun$ (blue) and the other parameters are fixed as $\Mbh=10^{2.5}\,\Msun,$ and $\beta=1.0$. The solid lines express the specific orbital energy distributions, and the dashed lines are the released nuclear specific energies if the compositions become pure $^{56}$Ni.
}
    \label{fig:epsilon_hist}
\end{figure}

For WDs with $\Mwd=1.2\,\Msun$,
tidal force redistributes the specific orbital energies
more effectively than the released nuclear energy does.
The energy spread due to the tidal force $\Delta \epsilon_t$ can be estimated as
\begin{align}
\Delta \epsilon_t&\sim\beta^n \frac{G \Mbh \Rwd} {R_t^{2}}\\
&\simeq1.2\times 10^{-3} c^2 \beta^n
\left (\frac{\Rwd}{10^9\,\textrm{cm}}\right )^{-1}
\left (\frac{\Mbh}{10^3\,\Msun}\right )^{1/3}
\left (\frac{\Mwd}{0.6\,\Msun}\right )^{2/3},
\label{eq:spread}
\end{align}
 with $n=2$ for a canonical model, while recent studies \citep{2013ApJ...767...25G,2013MNRAS.435.1809S} show that the value should be revised as $n=0$. 
Because heavier WDs have smaller radii (see Table~\ref{tab:Initwd}),
$\Delta\epsilon_t$ is larger for heavier WDs. 
If the nuclear reactions advance completely and the compositions become
almost pure $^{56}$Ni,
the released specific nuclear energies $\Delta\epsilon_{\textrm{nuc}}$ would be $1.7\times10^{-3},\,8.7\times10^{-4}$ and $6.8\times10^{-4}\,c^2$ with the same initial compositions as $\Mwd=0.2$, 0.6, and $1.2\,\Msun$ respectively. 
The effect of the nuclear reactions on the motion of the debris is less important
if $\Delta \epsilon_t \gg \Delta\epsilon_{\textrm{nuc}}$, and the amount of unbound debris does not
significantly increase. 
In Fig.~\ref{fig:epsilon_hist}, we show the specific energy distributions given by our simulations. 
The spreads of the distributions are consistent with Equation~(\ref{eq:spread}). 
We can see that $\Delta \epsilon_t \lesssim \Delta\epsilon_{\textrm{nuc}}$ is satisfied for
$\Mwd=0.2$ and $0.6\,\Msun$ cases. Thus, for those cases, debris where nuclear reactions
products IGEs or IMEs are mostly unbound. 
In contrast, for $\Mwd=1.2\,\Msun$ cases, $\Delta \epsilon_t \gtrsim \Delta\epsilon_{\textrm{nuc}}$ so that explosive nuclear reactions do not significantly increase the amount of unbound debris and a good fraction of IMEs/IGEs is bound.

\section{Resolution dependence}
\label{section:resolution}
\begin{figure}
	\centering
	\includegraphics[width=\hsize]{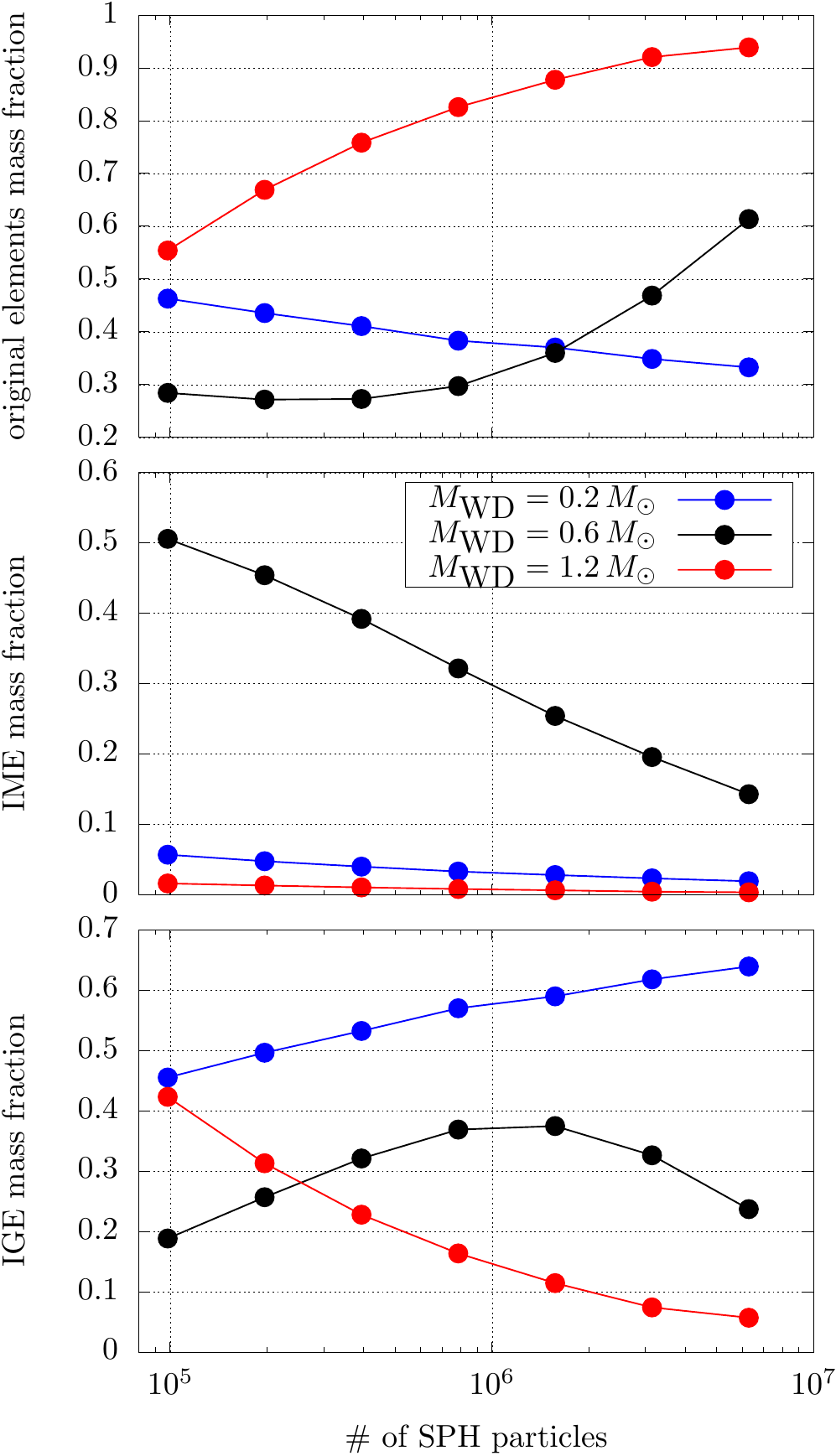}
	\caption{
Resolution dependence of nucleosynthesis. 
Here, the parameters are fixed as 
$\Mbh=10^3\,\Msun$ and $\beta=10$ for $\Mwd=0.2\,\Msun$, 
$\Mbh=10^3\,\Msun$ and $\beta=6$ for $\Mwd=0.6\,\Msun$, and
$\Mbh=10^{2.5}\,\Msun$ and $\beta=3.5$ for $\Mwd=1.2\,\Msun$. 
Note that the number of SPH particles we use in the parameter study is $\simeq 8\times10^5$.
}
    \label{fig:resolution}
\end{figure}
\begin{figure}
	\centering
	\includegraphics[width=\hsize]{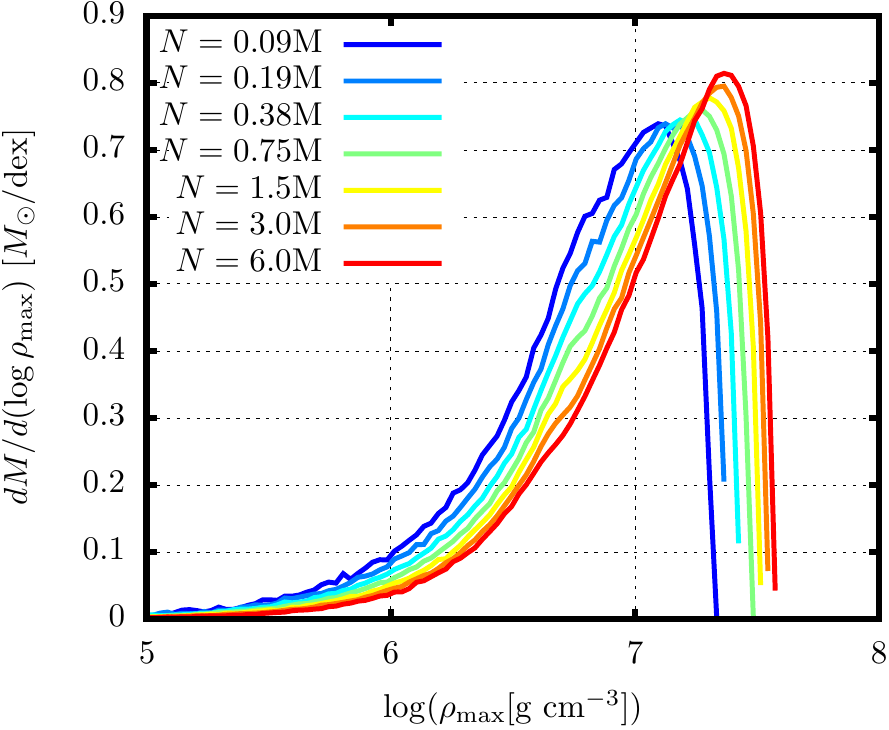}
	\caption{%
Resolution dependence of $\rhomax$. 
We show distributions of $\rhomax$ for the case of
$\Mwd=0.6\,\Msun$, $\Mbh=10^3\,\Msun$, and $\beta=5.0$ 
with different resolutions. 
We turn off nuclear reactions here. 
}
    \label{fig:rho_hist_resolution06}
\end{figure}
\begin{figure*}
\centering
\includegraphics[width=\hsize]{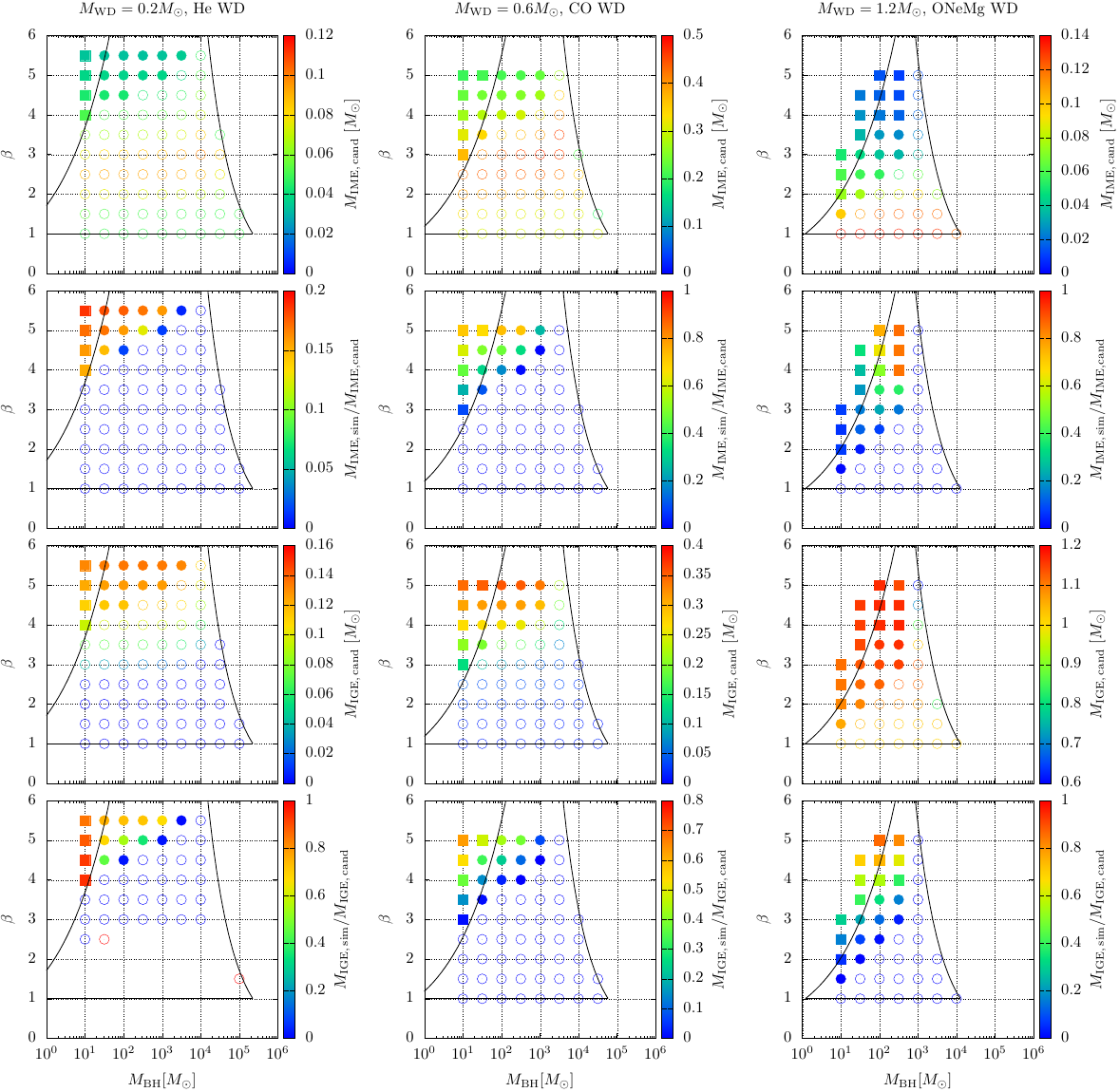}
\caption{%
Nucleosynthetic yields independent on resolution of calculations. 
The first and third rows show the IMEs/IGEs ``candidates'' masses, 
or $M_{\mathrm{IME, \,cand}}$/$M_{\mathrm{IGE, \,cand}}$, calculated from 
distributions of $\rhomax$ given by simulations uncoupled with nuclear reactions. 
The second and fourth rows show 
ratios of the masses of IMEs/IGEs synthesized in our former calculations 
coupled with nuclear reactions, 
or $M_{\mathrm{IME, \,sim}}$/$M_{\mathrm{IGE, \,sim}}$, 
to $M_{\mathrm{IME, \,cand}}$/$M_{\mathrm{IGE, \,cand}}$.
}
\label{fig:burned_fraction}
\end{figure*}
\citet{2017ApJ...839...81T} perform simulations of WD--BH TDEs 
where nuclear reactions ignite, using almost the same methods as ours.
Specifically, they take parameters as $\Mbh=500\,\Msun, \, \beta=5$ for
He/CO WD cases and $\Mbh=100\Msun, \, \beta=3$ for ONeMg WD cases.
They examine convergence of nucleosynthesis varying the
number of the SPH particles up to 25 million.
The results do not converge in all the cases.
They show that spurious heating due to low resolution ignite nuclear burning. 
They additionally perform 1D simulations with high resolution where
initial conditions are taken from three-dimensional (3D) SPH simulations uncoupled with
nuclear reactions.
In the 1D simulations, shock wave triggers detonation instead of spurious heating. 
However, the 1D initial conditions are not so accurate that we cannot precisely determine where a shock wave emerges.

We examine the resolution dependence of our nucleosynthesis results, which
is shown in Fig.~\ref{fig:resolution}. Here, we take the simulation
parameters such that $\beta$ takes the maximum value among the Type II
TDEs in order to avoid the effects of the early self-intersection and in
order to examine the effects of the failure to resolve the maximum compression point.
Fig.~\ref{fig:resolution} shows that our results do not converge,
although the degree of resolution dependence varies depending on the WD
masses. 
Our results are consistent with those of \citet{2017ApJ...839...81T}, in which different simulation parameters are used. 
The result shown in \figref{fig:resolution} renders our results of nucleosynthesis to remain uncertain.

In order to examine the validity of our results, we estimate the
nucleosynthetic yields in a different way and compare the 
both yields given in the two different ways.
To this end, we need a physical quantity that indicates 
nucleosynthetic yields, 
is not affected by the spurious heating, and thus is robust. 
Naively, the temperature and density when nuclear reactions occur
are key physical quantities associated with 
the nucleosynthesis (see \figref{fig:TRhotmax}).
However, the numerical heating critically affects $\Tmax$ and
$\rhotmax$, and we cannot use these values.

Instead, we additionally perform SPH simulations in which the nuclear
reactions are turned off, and record the maximum density of each SPH
particle during the simulation, $\rhomax$, to derive nucleosynthetic
yields. 
Most of the SPH particles have the maximum density $\rhomax$ when the 
tidal compression is the strongest; then the nuclear burning would be the
most violent if it happen.
We show the resolution dependence of $\rhomax$ in
\figref{fig:rho_hist_resolution06}, which shows that the distribution of $\rhomax$ indeed converges.

We estimate the elemental abundances as functions of
initial compositions and $\rhomax$ in 
\tabref{tab:burned_fraction}.
The table is based on the results of
\citet{2010A&A...514A..53F}, \citet{2013ApJ...771...14H}, and \citet{2015A&A...580A.118M}. 
\citet{2010A&A...514A..53F}, and \citet{2015A&A...580A.118M} estimate the
elemental abundances as functions of fuel density in detonations in Type Ia SNe by iterating hydrodynamical simulations 
of WD explosions and post-processing nuclear reaction
calculations.
\citet{2010A&A...514A..53F} derive corresponding functions for the helium and
carbon/oxygen 
detonation, while \citet{2015A&A...580A.118M} study detonation for initial
composition of $X(^{12}\mathrm{C})=0.03$, $X(^{16}\mathrm{O})=0.6$, 
and $X(^{20}\mathrm{Ne})=0.07$.
We apply these results for our CO and ONeMg WD cases. 
For He WD cases, we apply the results of \citet{2013ApJ...771...14H} and ours shown in
\figref{fig:TRhotmax}. 
Our approximation to the results is listed in \tabref{tab:burned_fraction}.
We calculate nucleosynthetic yields using
\tabref{tab:burned_fraction}, 
assuming that the whole WD experiences detonation and 
nuclear burning at the maximum compression point with the density
$\rhomax$. 

The results are shown in \figref{fig:burned_fraction}. 
Note that the masses of IMEs/IGEs obtained in this way are those of
IMEs/IGEs ``candidates'', or $M_{\mathrm{IME, \,cand}}$/$M_{\mathrm{IGE,
cand}}$, because we assume the whole WD burns.
If a part of the WD does not burn or produce IMEs or IGEs, the IMEs/IGEs masses are smaller than $M_{\mathrm{IME, \,cand}}$/$M_{\mathrm{IGE, \,cand}}$. 
In this sense, we can regard $M_{\mathrm{IME, \,cand}}$/$M_{\mathrm{IGE, \,cand}}$
as \textit{upper} limits of the IMEs/IGEs masses.
\figref{fig:burned_fraction} also shows the ratios of the masses of IMEs/IGEs synthesized in our
former simulations 
coupled with nuclear reactions, or $M_{\mathrm{IME, \,sim}}$/$M_{\mathrm{IGE, \,sim}}$,
to $M_{\mathrm{IME, \,cand}}$/$M_{\mathrm{IGE, \,cand}}$.
In all the cases, the ratios are smaller than unity, which means that 
$M_{\mathrm{IME, \,sim}} < M_{\mathrm{IME, \,cand}}$ and
$M_{\mathrm{IGE, \,sim}} < M_{\mathrm{IGE, \,cand}}$ are satisfied.
In many cases, the synthesized amounts are remarkably similar to the
results
shown in \figref{fig:fig_GN_tot_show}, especially in runs with large $\beta$ and small $\Mbh$.
Although there are cases where 
$M_{\mathrm{IME, \,cand}}$/$M_{\mathrm{IGE, \,cand}}$
are significantly larger than $M_{\mathrm{IME, \,sim}}$/$M_{\mathrm{IGE,
\,sim}}$, this is largely because of our assumption adopted here that
the whole WD is burned. 
Therefore, while there still remain uncertainties in the
nucleosynthesis calculations presented in \secref{section:results}, the
numerical resolution issue does not completely ruin our main conclusion that WD--BH TDEs produce a variety of events. 
The diversity is at least qualitatively represented by our series of hydrodynamics simulations.

The ``candidates'' masses are robust because of their convergence.
\figref{fig:burned_fraction} shows that they have a wide range,
potentially indicating that there would be a variety of the amounts of
IMEs/IGEs synthesized. 
The ratios $M_{\mathrm{IME, \,sim}}/M_{\mathrm{IME, \,cand}}$ and 
$M_{\mathrm{IGE, \,sim}}/M_{\mathrm{IGE, \,cand}}$
are still uncertain, 
and should be derived with extremely high-resolution studies, such as \citet{2017arXiv171105451T}. 
\citet{2017arXiv171105451T} perform 3D hydrodynamical simulations of He
WD--IMBH TDEs uncoupled with nuclear
reactions and multiple 1D high-resolution simulation coupled with
nuclear reactions for one parameter set. 
They investigate which part of the WD experiences detonation with
multiple 1D simulations, and give amounts of synthesized elements.
However, it is hard to perform such numerically expensive simulations for
various cases of WD--BH TDEs and to explore the variety of the amounts. 
\begin{table}
\centering
\caption{%
Nucleosynthetic yields as functions of initial compositions and
    $\rhomax$.
$X_{\mathrm{IME}}$, $X_{\mathrm{IGE}}$ are respectively 
mass fractions of IMEs/IGEs after nuclear burning.
}
\label{tab:burned_fraction}
\begin{tabular}{cccc}
\hline
Composition & $\rho_{\max}$ (g cm$^{-3}$)                         & $X_{\mathrm{IME}}$ & $X_{\mathrm{IGE}}$ \\ \hline
He          & $\rho_{\max} \leq 10^5$                         & 0                  & 0                  \\
            & $10^5 < \rho_{\max} \leq 5\times 10^5$          & 0.6                & 0                  \\
            & $5 \times10^5 < \rho_{\max}$                         & 0                  & 1.0                \\ \hline
CO          & $\rho_{\max} \leq 2 \times 10^5$                & 0                  & 0                  \\
            & $2 \times 10^5 < \rho_{\max} \leq 2\times 10^6$ & 0.4                & 0                  \\
            & $2 \times 10^6 < \rho_{\max} \leq 10^7$         & 0.9                & 0.1                \\
            & $10^7 < \rho_{\max}$                                 & 0                  & 1.0                \\ \hline
ONeMg       & $\rho_{\max} \leq 2 \times 10^5$                & 0                  & 0                  \\
            & $2 \times 10^5 < \rho_{\max} \leq 2\times 10^6$ & 0.15               & 0                  \\
            & $2 \times 10^6 < \rho_{\max} \leq 10^7$ & 0.8                & 0                  \\
            & $10^7 < \rho_{\max}$                                 & 0                  & 1.0                \\ \hline
\end{tabular}%
\end{table}
\section{Conclusions}
\label{section:conclusions}
We have performed a large number of simulations of WD--BH TDEs with a wide variety of
the WD mass, BH mass and the initial orbital parameters. 
We find that, when explosive nuclear reactions occur, a significant mass fraction of a WD,
up to a few tens percent, is converted into IGEs.
Most of them are gravitationally unbound, and radioactive decay of $^{56}$Ni in the unbound ejecta
would cause a transient event
similar to Type I SNe \citep{2016ApJ...819....3M}.
The released nuclear energy increases the fraction of unbound debris,
but the effect is less important for heavier WDs because the distribution of specific
energies of the debris is determined by the tidal force rather than nuclear reactions. 

There still remain uncertainties in our nucleosynthesis results due to
inadequate resolution in our hydrodynamical simulations coupled with nuclear
reactions.
In many cases, however, the synthesized elements masses are quite similar to upper limits of them obtained by a robust method with which we can avoid the uncertainties.
The results derived in the two different ways consistently show that WD--BH TDEs produce a wide variety of nucleosynthetic yields.

\begin{figure}
	\centering
    \includegraphics[width=\hsize]{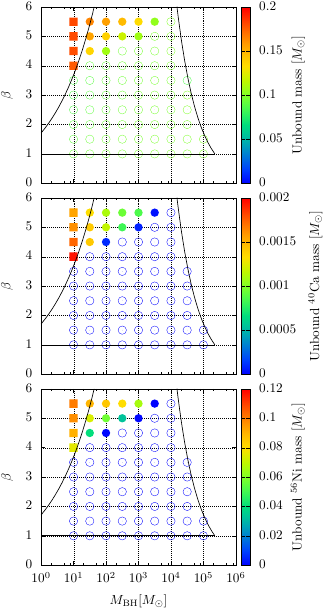}
	\caption{Properties related to calcium-rich gap transients. We show the $0.2\,\Msun$ WD cases here.
}
    \label{fig:calcium-rich}
\end{figure}
\citet{2015MNRAS.450.4198S} argue that a TDE of a light WD mainly composed of $^4$He
by an IMBH can be a progenitor of calcium-rich gap transients. \citet{2012ApJ...755..161K}
show properties of the transients: they are similar to Type I SNe, but are faint, with high velocities,
very large calcium abundances, and faster evolution than average Type Ia SNe.
Calcium-rich gap transients are found typically in the outskirts of known galaxies. 
We can derive the elemental abundances after TDEs where a $0.2\,\Msun$ WD is disrupted. 
Fig.~\ref{fig:calcium-rich} shows the masses of $^{40}$Ca and $^{56}$Ni in unbound ejecta.
$^{40}$Ca accounts for a tiny fraction of the ejecta  ($\lesssim$1~\%), while $^{56}$Ni does a large fraction ($\simeq$50~\%) in the Type II and III TDEs. 
The large masses of $^{56}$Ni ($\gtrsim0.04\,\Msun$) do not match with the estimated $^{56}$Ni masses in observed calcium-rich gap transients, such as 0.003~$\Msun$ for SN2005E \citep{2010Natur.465..322P}, 0.016~$\Msun$ for PTF10iuv \citep{2012ApJ...755..161K}, and 0.005-0.010$~\Msun$ for SN2012hn \citep{2014MNRAS.437.1519V}. 
We do not find WD--BH TDEs that can be a progenitor of calcium--rich gap transients. 
This is because the WD is so strongly compressed by the tidal force that $\rhotmax$
reaches $\gtrsim10^6$~g~cm$^{-3}$, the explosive nuclear reactions leave a small amount
of IMEs ($\lesssim0.01\,\Msun$) and a large amount of IGEs ($\sim0.1\,\Msun$). 

In the cases of Type III TDEs where the early self-intersection occurs,
extremely super-Eddington accretion rates would be realized. 
In such situations, there would be relativistic winds and/or jets driven by
radiation pressure, and these events could be so-called jetted TDEs 
\citep{2009MNRAS.400.2070S,2014ApJ...781...82C,2014ApJ...784...87S,2015ApJ...809....2C,2017MNRAS.471.1141L}.
An observer along with the jet axis would see beamed emission resembling
ultra-long $\gamma$-ray bursts that would be observable even if the events
are very distant.
For example, \citet{2016ApJ...819....3M} estimate that events with the jet luminosity of $10^{49}\,\mathrm{erg}\,\mathrm{s}^{-1}$ 
would be observable up to the redshift of 2.45 with Swift Burst Alert Telescope.

We have explored the diversity WD--BH TDEs caused by the combination of
the physical parameters, $\Mwd$, $\Mbh$, and $\beta$.
The variety of TDEs may show a wealth of phenomena observationally. 
\citet{2016ApJ...819....3M} perform radiative transfer calculations using the
results of hydrodynamical simulations of \citet{2009ApJ...695..404R}. 
They find that a single TDE can show different observational signatures
depending on the viewing angle. 
We plan to perform detailed radiative transfer calculations to
derive the spectral energy distributions and light curves in our future work.

\section*{Acknowledgments}
K.K. thanks Kazumi Kashiyama, Conor M. B. Omand, and Toshikazu Shigeyama for helpful discussions and advices.
We also thank anonymous referees for their helpful comments and
suggestions.
K.K. is supported by Advanced Leading Graduate Course for Photon Science. 
A.T. is supported by Grants-in-Aid for Scientific Research (16K17656) from 
the Japan Society for the Promotion of Science.
K.K. and N.Y. acknowledge financial support from JST CREST Grant Number JPMJCR1414.
Numerical calculations in this work were carried out on Cray XC40 at the Yukawa Institute Computer Facility and on Cray XC30 at Center for Computational Astrophysics, National Astronomical Observatory of Japan.



\bibliographystyle{mnras}
\bibliography{tde} 




%
%


\bsp	
\label{lastpage}
\end{document}